# A Minimalistic Approach to Predict and Understand the Relation of App Usage with Students' Academic Performances

MD SABBIR AHMED, North South University, Bangladesh
RAHAT JAHANGIR RONY, Cardiff University, UK
MOHAMMAD ABDUL HADI, University of British Columbia, Canada
EKRAM HOSSAIN, University of Rochester, USA
NOVA AHMED, North South University, Bangladesh

Due to usage of self-reported data which may contain biasness, the existing studies may not unveil the exact relation between academic grades and app categories such as *Video*. Additionally, the existing systems' requirement for data of prolonged period to predict grades may not facilitate early intervention to improve it. Thus, we presented an app that retrieves past 7 days' actual app usage data within a second (Mean=0.31s, SD=1.1s). Our analysis on 124 Bangladeshi students' real-time data demonstrates app usage sessions have a significant (p<0.05) negative association with CGPA. However, the *Productivity* and *Books* categories have a significant positive association whereas *Video* has a significant negative association. Moreover, the high and low CGPA holders have significantly different app usage behavior. Leveraging only the instantly accessed data, our machine learning model predicts CGPA within ±0.36 of the actual CGPA. We discuss the design implications that can be potential for students to improve grades.

## 1  INTRODUCTION

In recent years, smartphone use increased much among young people [2, 38]. Researchers found a negative relation of academic performance with smartphone usage duration [5, 6] as well as usage duration on a single app (e.g., KakaoTalk [11], Facebook [27]) which uncovers the necessity to understand smartphone's relation in depth. In Bangladesh, 86% of university students use smartphones [16] which also shows the requirement of in-depth exploration.

The research community has explored the topic with great interest. There is contemporary research regarding the relation of academic performance with *Social Media* [1, 4, 26, 27] and other apps [11, 13, 22, 23, 24, 25, 34, 50]. Through exploration of subjective data, some of these studies found *Productivity* apps such as *Notes* [22], and *Books* apps such as *Dictionary* [23] help students in learning. Despite much popularity of the *Books & Reference, Productivity,* and *Video Players & Editors* [37] categories, to our best knowledge, none of the previous studies used objective data of these categories to explore the students' usage behavior and its relation. But exploration through objective data has potential as self-reported app usage data cannot capture the actual behavior [5, 11]. Moreover, parents' concerns about the smartphone usage of their children [49] and also an exploration of previous studies [5, 6, 11, 13, 20, 21, 72] regarding the negative relation of smartphones with academic performance shows the significance of the investigation. Beside these, to predict academic performance, previous studies used different data such as behavioral [15, 33, 81, 89], internet usage [73], and e-learning [43, 74] where long-term data is required (e.g., 10 weeks [15], 2 years [33], 3 years [89]). Therefore, in this study, we fill these gaps regarding the exploration of objective data and the development of a system which does not need long-term data. We aim to find out the relation of smartphones with academic performance (Section 5.1 and Section 5.3), focusing on the *Books & Reference, Productivity, Video Players & Editors,* and *Social Media* app categories (Section 5.2 and Section 5.3). We also present the detailed app usage behavior of the high and low academic performers (Section 5.4). In addition, we aim to predict academic performance instantly through a minimalistic and unobtrusive way (Section 5.5).

In this study, we developed a tool which can retrieve the past 7 days' actual app usage data. To estimate the required time for data retrieval, we tested the tool by retrieving the past 7 days' data 10,000 times from 20 students' personal smartphones which were of 19 different models. In addition, to assess the impact of app usage on academic results and to develop predictive models, we collect 7 days' actual app usage data from 124 Bangladeshi students. We explore the correlation of academic results with aggregated, diurnal (a type of daily activity pattern [98]) usage data of students' used 27 app categories. Later, we investigate the differences in academic results between the high and low users. In addition, using an unsupervised machine learning (ML) method, we cluster the students who have similar smartphone usage behavior. After that, we investigate whether there is any variation in academic results of the high and low users who have similar smartphone usage behavior but different usage behavior regarding the app categories. Besides these, to explore the smartphone usage behavior of the high and low academic result holders, we do a comparative study. Finally, after finding the best estimator through a 5-fold cross-validation, we use our ML models in predicting academic results on the basis of app usage data only.

We find on average, our developed tool retrieves the past 7 days' app usage data by a second (Mean=0.31 second, SD=1.1 second). Our statistical analysis shows that the relation of app usage with academic results (CGPA: Cumulative Grade Point Average) is not one-dimensional - it varies based on several parameters. We find that apps usage sessions have a statistically significant negative relation with CGPA. In addition, exploring 27 app categories, we find that *Productivity* and *Books & Reference* app categories have a significant ($p<0.05$) positive relation whereas *Video Players & Editors* category has a significant ($p<0.05$) negative relation with CGPA. However, an app category such as *Social Media* does not show any statistically significant relation. Moreover, exploring the difference between high and low performers, our findings show that the high and low CGPA holders have significantly different usage behavior in *Productivity* and *Video Players & Editors* app categories. Beside these, using only app usage data as features for ML models, we find that students' CGPA can be predicted accurately. We find that the predicted CGPA of the KNN algorithm-based model is within ±0.36.

Our study makes several research contributions in computing education and HCI areas.

- Firstly, from only the instantly accessed 7 days' app usage data, our ML model predicts academic performance accurately, and to the best of our knowledge, our presented approach is faster, minimalistic, and unobtrusive than any other existing approach which may play a role in low resource settings.

- Secondly, as far as we know, using objective data, we are first to present a detailed analysis regarding *Books*, *Productivity*, and *Video* apps revealing *Books* and *Productivity* can be supportive to improve CGPA.
- Thirdly, we present the high and low CGPA holders' app usage behavioral differences in data such as duration, launch, and number of apps presenting the usability of these data to categorize varying CGPA holders.
- Finally, based on our findings, we conclude this paper with design implications which can facilitate the students to concentrate more and take early intervention to improve academic performance.

## 2 RELATED WORK

### 2.1 Relation of Smartphone Use with Academic Performance

Scholarly studies [5, 11, 13, 20, 21, 45, 76, 77] used smartphone usage data to investigate its relation with academic performance and most of these studies found a negative relation. A year-long study [12] conducted over 24 students found the smartphone as detrimental to educational goals. Smartphone usage for specific purposes such as video gaming, internet searching, etc. also showed a negative relation with GPA [13]. Considering specific contexts like a classroom, researchers studied this relation between smartphone usage and academic results. During class time, students use their phones for more than one-fourth of their effective class duration and they are distracted by the phone once every 3-4 minutes [11]. Researchers [76] found a negative association between smartphone usage during class time and academic results. In fact, the negative impact on academic results becomes almost two times higher if phone usage happens during class time [5]. Lepp et al. [21] discussed that the high frequent users may focus less on academic pursuits which can cause a negative relation with GPA. However, during class time, students who willfully neglect notifications, use smartphones less and have good academic performance [76]. Considering specific as well as general settings, though many studies investigated the impact of smartphones in the context of developed countries, there is a very little study that used actual data to investigate such impact in developing regions. Understanding the students' smartphone usage behavior and its relation with academic results is an active research area where a study in the context of developing regions can supplement the existing studies. In addition, as there is a variation of smartphone usage among the countries [31], study in the developing context can be the potential to develop a generalizable system to improve students' academic results. In the context of Bangladesh, a previous study [6] found a lower CGPA of high smartphone users. However, their study is limited by only visualization and does not have any statistical evidence behind such a negative relation.

#### 2.1.1 Relation of Books and Productivity apps with academic performance

Prior studies found there are several apps that help students in the learning process. For instance, researchers found mobile dictionaries [18], apps such as *Learn English Grammar* [25] help in learning. Moreover, mobile dictionary helps in vocabulary learning [23] and students refer to mobile dictionaries for correct pronunciation, spelling [23], portability, and accessibility issues [23]. Besides this, the findings of another prior study [24] revealed that students find comics as fun, and leisure that can reduce stress. Apart from the *Books & Reference* category-related apps, researchers talk about apps of the *Productivity* category which also helps the students. For instance, Wai et al. [22] found that apps such as *Notes*, *Document Viewer*, etc. are useful for learning purposes. The main limitation of these studies is the use of self-reported data which can provide biased findings, and also the app usage data such as duration, duration per launch, etc. is unexplored to investigate such a relation which can be more useful in understanding the impact and also in developing a personalized model.

### 2.1.2　Relation of Video apps with academic performance

Little research has been conducted regarding the relation of the *Video Players & Editors* app category. Uzun et al. [13] surveyed 631 university students and found that entertaining apps are negatively correlated with GPA. Another survey conducted by over 2000 students, found *YouTube* causes a bigger distraction than *Social Media* apps [34]. Moghavvemi et al. [8] found that around 91% of students use *YouTube* for entertainment purposes and 55% of students use it for academic learning. Previous studies [13, 34] regarding video-related apps, used self-reported data to investigate the relation. But self-reported data cannot capture the actual smartphone usage behavior [5, 11].

### 2.1.3　Relation of Social Media apps with academic performance

There are different apps such as *Facebook*, *Instagram*, etc. in the *Social Media* category. Nevertheless, most previous studies [1, 3, 4, 9, 10, 27] were conducted regarding the relation of *Facebook* where other *Social Media* apps less explored. Junco [27] found that time spent on *Facebook* is significantly negatively related to academic performance. Several other studies [13, 30] also found such a negative relation. On the other hand, Wang et al. [1] did not find any significant difference in CGPA between the high and low *Facebook* checkers. Another study [26] found continual checkers have a lack of control in *Social Media* usage, though, their GPA is not different from the infrequent checkers. The relation of *Facebook* with academic performance was also found to vary by device [10], the purpose of using *Facebook* [3], *Facebook* usage data [4], and continent (Europe and North America) [35]. In *Social Media*, though some studies used actual data, these studies focused on a few apps (e.g., *Facebook* [1], *Instagram* [47]) only or did not provide the detailed analysis (e.g., [26]) regarding the relation of *Social Media* with academic performance.

## 2.2　Prediction of Academic Performance Through Different Types of Data

In some studies, researchers used behavioral [15, 33, 81, 89] data such as *Facebook* activity data [33] to classify the performers of 3 different levels based on GPA, and orderliness data [81, 89] to predict academic performance. Using the students' 10 weeks long smartphone sensed and also self-reported data, Wang et al. [15] predicted CGPA which is within ±0.17 of the actual CGPA. Some other studies [43, 44, 74] used educational data where Waheed et al. [74] used 9 months' data of virtual learning environments, and Leppänen et al. [43] used HTML element level data of learning material from 7 weeks long course. There are some other studies conducted to predict students' performance using self-reported data such as family tutoring, school tutoring [93], demographic characteristics, study, and family related data [17]. However, as aforementioned, the self-reported data can have biasness [5, 11] and also the requirement of self-reported data can make the system intrusive which can be disrupting to a student. On the other hand, the main research gap in the existing objective data-based studies is the requirement of long-term data (e.g., 3 years [81], 7 weeks [43]) where instantly accessed data could play an important role for early intervention to improve academic results.

## 2.3　Design to Improve Academic Performance Through Intervention in Smartphone Usage

With great interest, researchers investigated design approaches to mitigate meaningless experiences in app usage [59]. Studies also explored how intervention can discourage app usage [69], how blocking systems can be productive [85], and what can encourage habit formation [54]. Researchers found the group-based system performs better in minimizing the users' phone usage [71, 84, 97], in reducing distraction, and also in focusing study more [84]. In group mode, one is influenced by the other in limiting her/his own phone usage through comparison with others [80] as well as by peer pressure [84]. Beside these, goal-based systems (e.g., self-defined duration [58]) were found to limit smartphone usage [58] as well as specific app usage such as *Facebook* [61, 83]. Some other studies [79, 80, 84] explored understanding contexts to design systems. Through designing a software-based intervention system for the classroom, Kim et al. [80] discussed limiting phone usage along with retaining students' autonomy. However, designing systems considering the relation of app categories with academic performance is less explored. Though few studies [11, 76] suggested designs based on exploring that relation, those research were conducted in limited settings such as classrooms which may not be representative of the whole day. As the usage behavior varies by the app category [7], all categories may not have identical relation with the grades. Thus, taking into account the relations with app categories can facilitate having a better system to improve the students' grades.

## 3  DATA COLLECTION TOOL

We developed a tool (app) for data collection. In a previous study [15], researchers used data retrieved from smartphone sensors (e.g., GPS) and they needed additional preprocessing which can be computationally and power expensive. For example, they needed classifiers to classify the activities, detect conversation, and sleep [32]. But we use only app usage data and our tool does not rely on sensors to retrieve data. Also, our tool does not need to run in the background.

### 3.1  Development and Testing the Data Collection Tool

We considered only the Android users as 95% of phone users from Bangladesh use this platform [29]. Using different functions of the Java class *UsageStatsManager* [53], we retrieved the past 7 days' foreground and background app usage events instantly (Section 3.2) where an event is an action of user to open (foreground event) or close (background event) an app. We explored and did many experiments to find out the ways for retrieving the past days' data accurately. However, all procedures were not accurate. One of those approaches was exploring *queryUsageStats (int intervalType, long beginTime, long endTime)* [111]. By testing in some ways such as manually calculating the app usage duration, we found though the retrieved apps' package name and app name are accurate, the retrieved app usage duration was not accurate. One of the main drawbacks of this approach is *queryUsageStats()* cannot retrieve the raw events data (e.g., when an app was opened and closed). Instead, the function retrieves the accumulated (e.g., past 7 days) data at a time which impeded the way to extracting informative features (e.g., engage session [67]). On the other hand, we found *queryEvents (long beginTime, long endTime)* function [53] working as accurate when we retrieved[1] the data by dividing the past 7 days into 3 segments. Additionally, due to having a higher number of events in a smartphone, our app became slower which we resolved by creating threads. Our developed app needed two permissions: 1) camera permission to scan students' ID cards; 2) usage access permission to retrieve app usage data. In our app, there was a barcode scanner for scanning students' ID cards to ensure the entry of actual ID numbers as depending on that students' CGPA will be retrieved from the database.

We tested our tool [114] in different ways. At first, we calculated the app usage data such as duration manually. Then, we compared this data with the retrieved app usage data of our tool. We also compared the retrieved data of our tool with such apps [51, 52] available in the Google Play store which need to run in the background to get the app usage data [51, 52]. In each step, our data collection tool calculated the past 7 days' app usage data accurately.

---

[1] Core codes to retrieve data is available in our project (DOI: 10.17605/OSF.IO/923HM) and can be accessed at https://osf.io/7trk4 which is licensed under Apache License 2.0

## 3.2 Performance of the Tool in Retrieving Data

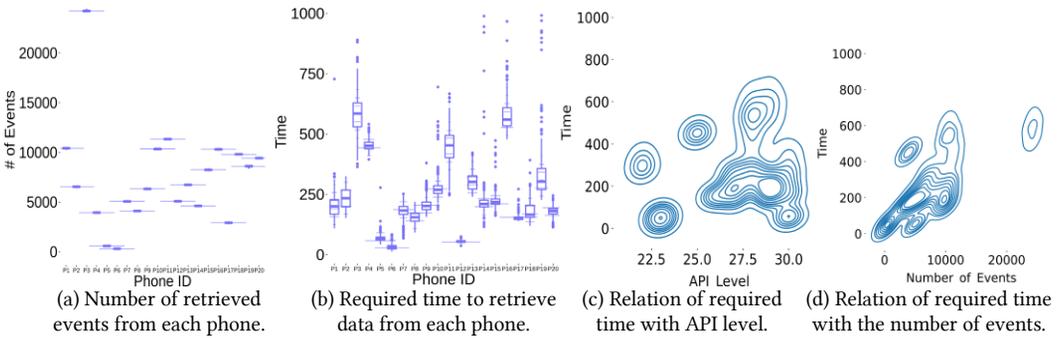

(a) Number of retrieved events from each phone.   (b) Required time to retrieve data from each phone.   (c) Relation of required time with API level.   (d) Relation of required time with the number of events.

Figure 1: Performance of the tool. (a) Number of retrieved foreground and background events, (b) Required time in retrieving data, KDE (Kernel density estimation) shows the relation of time with (c) API level and (d) number of foreground and background events.

We estimated the time required to retrieve the foreground and background events from smartphones through testing our tool on 20 smartphones from which the tool [114] retrieved data repetitively for more precise estimation. These phones had different configurations. From each phone, our tool retrieved the past 7 days' app usage data 500 times. In total, we calculated the required time 10,000 times. On average, the number of retrieved events was 7,447.61 (SD=4986.62) (Figure 1(a)) and the mean required time was 307.94 milliseconds (ms) (SD=1103.91 ms) (Figure 1(b)). Among 10,000 instances, only 97 instances were above 1 second. To understand the factor that can impact the time in data retrieval, we explored the correlation (the method is in Section 4.3.1) of required time with 20 phones' API (Application Programming Interface) level and the number of retrieved events. We found no significant correlation with the Android API level and required time ($r_s$= 0.18, p=0.44). However, we found a significant positive relation between the events' number and the required time to retrieve data ($r_s$=.56, p=.0096). After that to estimate the plausible number of events in the students' phones, we used 121 students' past 7 days' event usage data (Section 4.2). On average, there were 8,135.47 events (SD=3984.11). Among our data retrieval 10,000 times, there were 4,500 instances where the number of retrieved events was more than 8,000 and to retrieve these events, our app took 430.31 ms (SD=1596.455 ms) which says on average, our app can retrieve the past 7 days' app usage data within a second.

## 4 METHODOLOGY

### 4.1 Participants' Consent and Data Collection Procedure

We collected data from 124 undergraduate students who were from 4 different academic years and 3 departments (Computer Science & Engineering (CSE), Electrical and Electronic Engineering, and Bachelor of Business Administration) of a university situated in Dhaka, Bangladesh. We selected the participants using a snowball sampling method [46] and data was collected before the Spring-2019 semester final exam. Participants were given the consent form and they voluntarily donated data through our app. To quantify students' academic performance, we used cumulative grade point average (CGPA) as it is positively connected to the percentage of class attendance, semester GPA [55], and academic achievement [78], and is widely used [1, 6, 15, 32] in HCI studies to measure academic performance. We collected the actual CGPA from the university officials with permission from participating students. In the participants' university, the CGPA range is 0.0 to 4.0 and it is mostly based on scores of exams such as class tests, mid-term, and final exams.

## 4.2 Data Preprocessing

Among 124 participants, we analyzed session data, and diurnal usage data of 121 participants due to the unavailability of the relevant data in the database in case of 3 students. This problem can be due to network problems as the internet connection was unstable despite our creation of 3 mobile hotspots during the data collection period. Except for those 3 students, the total number of foreground and background events of 7 days' app usage was 984,392 (Min.=1128, Median=7446.0, Mean=8135.47, SD=3984.11, Max.=20376). Apart from calculating the duration per launch and duration per app, we extracted the following app usage data to understand the impact of app usage and also to develop models.

*Frequency of launch:* We calculated the frequency of launch by counting the number of foreground events of the apps.

*Usage duration:* App usage duration was calculated by subtracting the background event's time from the foreground event's time. In the usage duration of an app category, we aggregated all such sessions that occurred in apps of that category.

*Number of apps:* We calculated the number of used unique apps by counting the number of package names of the apps which were used instead of the app names as two apps may have the same name.

*Diurnal usage data:* We considered the period of a day for analysis. A prior study [14] presented that smartphone usage can vary with the change of place. As students need to move several places in a day (e.g., university, home), it might change their usage behavior as well. Thus, inspired by prior research [7, 40] and we categorized a day into four equal periods: Night: 12:01 AM - 6:00 AM; Morning: 6:01 AM - 12:00 PM; Afternoon: 12:01 PM - 6:00 PM; Evening: 6:01 PM - 12:00 AM. In Bangladesh, the evening is considered students' study time.

*App usage sessions:* To understand whether the sessions of app usage relate with academic performance, we count the number of sessions. The Java function (Section 3.1) which was used to collect 7-days' app usage data does not provide the time when the phone was locked and unlocked and thus, we followed previous studies. Wang and Mark [1] used the median break length (the duration between two consecutive Facebook usage) of 40 seconds, some studies used a threshold of 60 seconds [61, 62] and 30 seconds [63, 64, 65] to define the sequence of app usage into a single session. However, Berkel et al. [66] found 30 seconds threshold can be less accurate and they suggested using 45 seconds. Therefore, we group the app usage into a single session when there is no more than 45 seconds gap between the last used and the newly opened app. Meanwhile, Banovic et al. [67] presented three types of sessions and found users' different tasks with varying types of sessions. Hence, different types of sessions may create different impacts on academic performance. Based on this assumption, we categorize the sessions into three different types.

- *Micro session:* We count a micro session if a participant spends a maximum of 15 seconds in a session [68].
- *Review session:* Previous study [67] defines a review session if it lasts 60 seconds. Due to exploring micro-sessions, we count a review session if one spends more than 15 seconds but less than or equal to 60 seconds.
- *Engage session:* Following Banovic et al. [67], we count an engage session if the participants spend more than 60 seconds using the apps of a smartphone. In this type of session, users often interact with multiple apps [67].

*Data of the app categories:* In 7 days, 124 students used 884 different apps. We categorized these apps into 27 different categories (details are in Table 3 as presented on page 11). We took several steps during the app categorization. At first, we retrieved the developers' preferred category available in Google Play Store by using the package name and a Java HTML parser. However, students used several apps which were not available in that app store. We understood the features of those apps by exploring other app stores (e.g., apkmirror.com) and developers' websites. We categorized the apps following the previous studies [7, 65] and through a discussion with 3 students of the CSE department. For the apps where there was disagreement among the categorizers, we discussed with 2 more students and used the majority rule to select a category. Due to having different numbers of users in different app categories, the sample size varied by category as presented in the findings section.

## 4.3 Data Analysis

We did inferential statistical analyses to explore relation between app usage data and CGPA. We developed models to predict grades based on our instantly retrieved data. Our codes[2],[3] are available as open-source under Apache License 2.0.

### 4.3.1 Statistical analysis

**Correlation analysis.** Though correlation does not present causal relation, we were inspired by the previous studies [15, 32, 88, 90] of the HCI community to understand the association. Using *scipy* library [115], we used the Pearson (r) formula when both app usage data and CGPA were normally distributed and outliers free. Otherwise, we used the Spearman formula ($r_s$) which is nonparametric. To detect the outliers, we used the Z score method setting 3 as the threshold [87].

**Comparative analysis.** Having a relation between two variables cannot ensure the difference between two groups. For instance, having a significant positive correlation of CGPA with an app category x does not ensure that high users of x will have statistically significantly higher CGPA than the low users. Hence, to understand whether different usage data of smartphones significantly relate with academic performance, we also did a comparison between different groups of students. We used either the T-Test or the Mann-Whitney U test depending on the satisfaction of the assumptions. When the data were normally distributed and both of the groups had equal variance, we did the Standard T-test. In data having an unequal variance, we did Welch's T-test. In all other cases, we did a non-parametric Mann-Whitney U Test. As multiple comparisons can produce false-positive results (type I error), we adjusted p values using false discovery rate approach [42]. To compare the students, we divided them in the following ways:

- ***High and Low CGPA Holders:*** To explore the difference between the high and low CGPA holders, we divided the participants following previous studies conducted in Bangladesh [48] and the USA [15]. Participants having a CGPA of at least 3.50 and below 3.0 out of a CGPA of 4.0 were categorized as high and low CGPA holders respectively.
- ***Higher and Low Users:*** The top one-third and bottom one-third percentile on the basis of app usage data were considered as high and low users respectively [1, 26, 47].

### 4.3.2 Cluster analysis

To group the students who use smartphones for a similar amount, we have used the DBSCAN (Density-Based Spatial Clustering of Applications with Noise) clustering algorithm which is non-parametric in nature. The main motivation behind choosing this algorithm is that DBSCAN can automatically select the number of clusters which is not possible by clustering algorithms such as K-Means. Moreover, our participants' total smartphone usage duration and frequency of launch were not normally distributed which also motivated us to use this clustering algorithm.

### 4.3.3 Prediction of academic performance

**Feature extraction and selection:** From the foreground and background app usage events, we extracted 720 app usage behavioral markers which have been used as features of the predictive models: (27 app categories * 5 dur_launch_app_data * 5 time_periods) + (1 total smartphone usage data * (5 dur_launch_app_data + 4 session_data ) * 5 time_periods) where dur_launch_app_data = {duration, frequency of launch, # of apps, duration per app, duration per launch}; session_data = {total number of session, micro session, review session, engage session}; time_periods = {night, morning, afternoon, evening, whole day}. Before the development of the predictive model, we selected the important features using the least absolute shrinkage and selection operator (Lasso) algorithm which uses $l_1$ regularization and makes the less important features to 0. In our study, as there were more features than the number of participants, we used the Elastic net algorithm also as suggested by Géron [95]. Some previous studies (e.g., [70]) used the correlation formula for feature selection. Thus, apart from these approaches, we also select features using another strategy: select only the features which show a significant correlation ($p < 0.05$) with academic performance.

---

[2] Codes for processing the raw app usage events: https://osf.io/sxjp6
[3] Codes for statistical analysis and ML model development: https://osf.io/5xy6c

**Development of the predictive models:** To develop predictive models, we used a diverse set of algorithms: K-nearest neighbour (KNN), Support vector regression (SVR), Gradient boosting regression (GBR), Elastic net, Random forest, Lasso, Multilayer perceptron (MLP), Extreme gradient boosting (XGBoosting) algorithms [95] (brief description in Table S.1 of Supplementary). Some algorithms such as SVR depend on the distance between the data points. Before using those algorithms, we scale the features so that one feature does not get more importance due to having larger values in different units. In building machine learning (ML) models, generally, 80:20 ratios are used where 80% of data are kept for training and 20% of data are kept for testing purposes. There are studies [43, 60] which used 70:30 ratios. If we use 80:20 ratios, only 24 participants' data remain for testing. Therefore, we use a 70:30 ratio to keep a good number of samples for testing. Using the training dataset, we do a grid search with 5-fold cross-validation to find the best hyper parameters (Table S.2 of Supplementary) and the best estimator for each of the regression models presented in Figure 2. Finally, we group the best (in terms of average Mean Absolute Error (MAE) value in 5-fold cross-validation) 3 estimators using a meta estimator named *VotingRegression()* of *scikit-learn (sklearn)* [96] to build an ensemble model. To reduce the time complexity, we used 3 estimators and the main motivation behind building this ensemble model is to minimize the weakness of each estimator and use the concept wisdom of the crowd to get better performance as an ensemble method often performs better [95]. But if two algorithms use similar approaches to predict outcome values, they are more likely to make the same error. Therefore, during selecting the best 3 predictive models for voting regression, we focused on the MAE value and the diversity of the predictive algorithms. To build ML models, we used *sklearn* library [96].

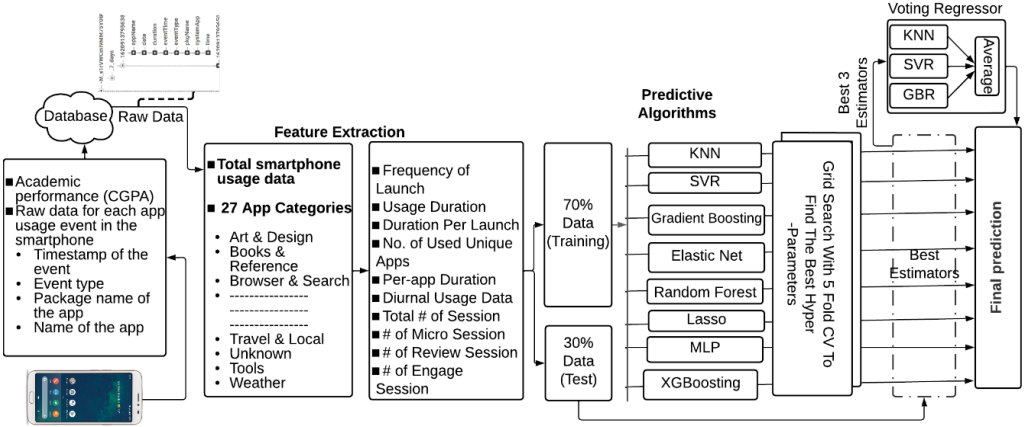

Figure 2: Pipeline for development of the ML models.

**Evaluation of the models:** We calculated the difference in MAE value of the training and validation set during the cross-validation to understand whether the model is overfitted. When we found that the difference is higher, we considered that model as the overfitted model. To solve the problem of overfitting, we changed the values in different ways. For instance, manually, we reduced the upper limit's value of different hyperparameters (e.g., max_depth) as it is one of the ways to solve the problem of overfitting [95]. After finalizing the model by finding the best set of hyperparameters from upper and lower limits of hyperparameters by grid search, we evaluate each estimator on the test dataset (30% data) which was not seen by the model during training. We also calculate the correlation of the predicted CGPA with the actual CGPA. Finally, we visualize this correlation to understand whether the models can predict variance in CGPA.

### 4.4 Research Ethics

This research was approved by the institutional center for research & development of a university. To make the participants aware of our research and the types of data we will be collecting, we arranged for a brief discussion session just before starting the data collection. All participants were provided a consent form where details regarding each of the collected data, future usage of the data, privacy, and data safety were explicitly written. Additionally, there was a privacy policy in the developer contact section of our app in the Google Play Store. Each participant consented to participate in the study. Additionally, except for participants' app usage data, our data collection tool did not collect any other data such as messages, photos, browsing history, etc. Besides, for participants' data privacy and safety, we took several steps such as participants' all data being encrypted in transit and stored participants' ID numbers as encrypted values Moreover, all data was stored in a secure storage which has restricted access only to the researchers of this project with two-factor authentication

## 5 FINDINGS

### 5.1 Relation of Smartphone Usage Data With CGPA

#### 5.1.1 Relation of aggregated smartphone usage data with CGPA

*The higher number of sessions relates to lower academic performance.* We find though total smartphone usage duration ($r_s$=-0.07, p=0.41), frequency of launch ($r_s$=-0.12, p=0.18), number of used apps ($r_s$= -0.098, p=0.28), duration per app ($r_s$= -0.001, p=0.99) have a negative relation with CGPA, the relations are not statistically significant (Table 1). But we find a significant negative association in the case of different app usage sessions. Our findings show that the total number of app usage sessions ($r_s$=-0.24, p=0.009), micro sessions ($r_s$=-0.198, p=0.03), and review sessions ($r_s$=-0.22, p=0.014) have a significant negative association with academic performance (Table 1). Moreover, the number of engage sessions also has a negative relation ($r_s$=-0.16, p=0.07) and the p-value is close to the significance level of 0.05.

Table 1: Correlation of CGPA with total smartphone usage data. Dark grey colored cells present a significant relation.

| Total Smartphone Usage Data | Sample Size (N) | Coef. (r.) | p | Total Smartphone Usage Data | Sample Size (N) | Coef. (r.) | P | Total Smartphone Usage Data | Sample Size (N) | Coef. (r.) | p |
|---|---|---|---|---|---|---|---|---|---|---|---|
| Duration | 124 | r.=-0.07 | 0.41 | # of Apps | 124 | r.= -0.098 | 0.28 | Micro Session | 121 | r.=-0.198 | 0.03 |
| Launch | 124 | r.=-0.12 | 0.18 | Duration/App | 124 | r.= -0.001 | 0.99 | Review Session | 121 | r.=-0.22 | 0.014 |
| Duration/Launch | 124 | r.=0.07 | 0.46 | Total  #  of | 121 | r.=-0.24 | 0.009 | Engage Session | 121 | r.=-0.16 | 0.07 |

#### 5.1.2 Relation of diurnal session data with CGPA

*App usage sessions in the afternoon and evening have negative relations with CGPA.* After finding negative relation in terms of whole days' data, we were interested to explore how these sessions relate with academic performance in different periods of a day. We find that the relation regarding app usage sessions at night and morning time is not statistically significant (p > 0.05) (Table 2).

However, in terms of sessions of afternoon and evening periods, there is a statistically significant negative relation of students' academic performance with the total number of sessions (afternoon: $r_s$=-0.21, p=0.01; evening: $r_s$=-0.21, p=0.01), review sessions (afternoon: $r_s$=-0.23, p=0.006; evening: $r_s$=-0.2, p=0.01), and micro sessions (afternoon: $r_s$=-0.22, p=0.008; evening: $r_s$=-0.19, p=0.02).

Table 2: Relation of CGPA with app usage session in four time periods. Dark grey-colored cells present significant results.

| Period | Total # of Session | | | Engage Session | | | Review Session | | | Micro Session | | |
|---|---|---|---|---|---|---|---|---|---|---|---|---|
| | Sample Size (N) | Coef. ($r_s$) | p | Sample Size (N) | Coef. ($r_s$) | p | Sample Size (N) | Coef. ($r_s$) | P | Sample Size (N) | Coef. ($r_s$) | p |
| Night | 121 | -0.08 | 0.21 | 120 | -0.06 | 0.53 | 106 | -0.02 | 0.42 | 109 | -0.15 | 0.055 |
| Morning | 121 | -0.14 | 0.07 | 121 | -0.08 | 0.39 | 121 | -0.13 | 0.08 | 121 | -0.12 | 0.097 |
| Afternoon | 121 | -0.21 | 0.01 | 121 | -0.1 | 0.3 | 120 | -0.23 | 0.006 | 120 | -0.22 | 0.008 |
| Evening | 121 | -0.21 | 0.01 | 121 | -0.15 | 0.11 | 120 | -0.2 | 0.01 | 120 | -0.19 | 0.02 |

## 5.2 Relation of App Category's Usage Data with CGPA

### 5.2.1 Relation in terms of aggregated usage data

After investigating the correlation of 27 different app categories with academic performance, we find that only some app categories (i.e., *Books & Reference, Productivity, Video Players & Editors*) show a significant correlation in terms of most of the usage data (Table 3). Though the number of used apps in the *Entertainment, Browser & Search, Health & Fitness* categories also show significant relation, these categories show no significant relation for other app usage data. In some app categories (e.g., *Art & Design*), the number of users (N) was very small. Thus, in the following sections, we present findings for the *Books & Reference, Productivity,* and *Video Players & Editors* categories. In addition, we present an analysis for the *Social Media* app category also as it has received much scholarly attention. To enhance the readability and keep the categories' names simple, we used *Books* instead of *Books & Reference* and *Video* instead of *Video Players & Editors.*

*Higher Books and Productivity apps use has a relation with higher CGPA.* In terms of usage duration, launch, and per-app duration of *Books* category apps, we find a statistically significant positive correlation with CGPA (Duration: $r_s$=0.24, p=0.026; Launch: $r_s$=0.22, p=0.045; Per-app duration: $r_s$=0.23, p=0.04) (Table 3). Similarly, in case of the *Productivity* category, we also find that students who spend more time ($r_s$=0.21, p=0.018), launch apps more ($r_s$=0.2, p=0.028), per-app duration is more ($r_s$=0.21, p=0.017), are more likely to have higher CGPA. Though duration per launch data does not show any significant relation with *Books* app category users' CGPA, it shows a significant relation with the CGPA of *Productivity* category's users ($r_s$=0.2, p=0.025) (Table 3).

*Lower Video use has a relation with higher CGPA but Social Media use does not have a relation with CGPA.* Unlike the *Books* and *Productivity* app categories, we find that the *Video* category relates with academic performance significantly negatively. The students who spend more time ($r_s$=-0.21, p=0.023), launch *Video* apps more ($r_s$=-0.21, p=0.022), and use more number of *Video* Apps ($r_s$=-0.19, p=0.040), are more likely to have lower CGPA (Table 3). However, investigating the correlation of CGPA with different usage data of *Social Media*, we find that none of the usage data of this category significantly correlate with academic performance (e.g., Duration: $r_s$=0.05, p=0.61).

Table 3: Correlation of CGPA with app category's usage data. Here, due to not having any variation in participants' used number of apps or due to having a very less number of users (N<=2), we set N/A. r presents Pearson correlation and others are Spearman correlation. Dark and light grey colored cells present significant and close to significant results respectively.

| App Category | Example Apps | % of Unique Apps | Sample Size (N) | Correlation of CGPA with usage data of app categories | | | | | | | | | |
| | | | | Usage Duration | | Frequency of Launch | | Duration per Launch | | No. of Used Apps | | Per-App Duration | |
| | | | | Coef. | p | Coef. | p | Coef. | p | Coef. | p | Coef. | p |
|---|---|---|---|---|---|---|---|---|---|---|---|---|---|
| Art & Design | Canva | 0.3 | 3 | -0.5 | 0.7 | 0.5 | 0.7 | -1.0 | 0 | N/A | N/A | -0.5 | 0.7 |
| Auto | Fuelio | 0.1 | 2 | N/A | N/A | N/A | N/A | N/A | N/A | N/A | N/A | N/A | N/A |
| Books | Translate | 4.8 | 83 | 0.24 | 0.03 | 0.22 | 0.045 | 0.16 | 0.2 | 0.12 | 0.3 | 0.23 | 0.04 |
| Browser | Chrome, | 2.3 | 124 | -0.14 | 0.1 | -0.09 | 0.3 | -0.1 | 0.3 | -0.2 | 0.03 | -0.07 | 0.4 |
| Business | Uber Driver | 0.7 | 7 | -0.25 | 0.6 | -0.27 | 0.6 | 0.11 | 0.8 | -0.2 | 0.7 | -0.32 | 0.5 |
| Communication | Contacts | 6.1 | 124 | 0.06 | 0.5 | -0.12 | 0.2 | 0.14 | 0.1 | -0.01 | 0.9 | 0.07 | 0.5 |
| Education | All C | 1 | 11 | r=0.29 | 0.4 | 0.03 | 0.9 | r=0.26 | 0.4 | 0.3 | 0.4 | 0.15 | 0.7 |
| Entertainment | WowBox, | 2.3 | 51 | 0.01 | 1 | -0.17 | 0.2 | 0.07 | 0.6 | -0.28 | 0.04 | 0.05 | 0.7 |
| Finance | bKash, Mi | 0.9 | 34 | 0.19 | 0.3 | 0.16 | 0.4 | 0.21 | 0.2 | 0.05 | 0.8 | 0.17 | 0.3 |
| Food & Drink | foodpanda | 0.1 | 4 | -0.8 | 0.2 | -0.63 | 0.4 | -1.0 | 0 | N/A | N/A | -0.8 | 0.2 |
| Games | Ludo Star, | 11.2 | 72 | 0.01 | 1 | 0.02 | 0.9 | 0.11 | 0.4 | -0.09 | 0.5 | 0.06 | 0.6 |
| Health | LG Health, | 1.2 | 12 | -0.52 | 0.1 | -0.31 | 0.3 | 0.04 | 0.9 | -0.64 | 0.02 | -0.28 | 0.4 |
| Lifestyle | Muslim+ | 0.8 | 11 | 0.07 | 0.8 | 0.09 | 0.8 | -0.03 | 0.9 | N/A | N/A | 0.07 | 0.8 |
| Medical | DIMS | 0.1 | 1 | N/A | N/A | N/A | N/A | N/A | N/A | N/A | N/A | N/A | N/A |
| Music & Audio | FM Radio | 5.4 | 91 | -0.04 | 0.7 | -0.03 | 0.8 | -0.09 | 0.4 | 0 | 1 | -0.07 | 0.5 |
| News & | BBC News | 0.3 | 9 | 0.18 | 0.7 | r=- | 0.2 | 0.29 | 0.4 | N/A | N/A | 0.18 | 0.7 |
| Personalization | Huawei Home | 6 | 124 | -0.08 | 0.4 | -0.05 | 0.6 | -0.1 | 0.3 | 0.05 | 0.5 | -0.13 | 0.2 |
| Photography | Camera, | 7.8 | 123 | -0.12 | 0.2 | -0.11 | 0.2 | -0.04 | 0.7 | -0.01 | 0.9 | -0.11 | 0.2 |
| Productivity | OfficeSuite | 4.8 | 124 | 0.21 | 0.02 | 0.2 | 0.03 | 0.2 | 0.03 | 0.13 | 0.1 | 0.21 | 0.02 |
| Shopping | AliExpress, | 1.1 | 21 | 0.12 | 0.6 | r=- | 0.9 | -0.0 | 1 | -0.22 | 0.3 | 0.17 | 0.5 |
| Social Media | LinkedIn | 1.5 | 118 | 0.05 | 0.6 | -0.02 | 0.8 | 0.07 | 0.4 | -0.12 | 0.2 | 0.16 | 0.1 |
| Sports | Cricbuzz | 1.1 | 16 | -0.18 | 0.5 | -0.17 | 0.5 | -0.19 | 0.5 | -0.34 | 0.2 | -0.14 | 0.6 |
| Tools | Orbot, | 32.1 | 124 | -0.13 | 0.2 | -0.05 | 0.6 | -0.03 | 0.7 | -0.05 | 0.6 | -0.1 | 0.3 |
| Travel | Booking.com, | 1.6 | 68 | 0.07 | 0.6 | 0.07 | 0.6 | 0.09 | 0.5 | -0.04 | 0.7 | 0.08 | 0.5 |
| Unknown | N/A | 0.8 | 8 | r=- | 1 | r=- | 0.2 | 0.07 | 0.9 | N/A | N/A | r=-0.03 | 0.9 |
| Video | YouTube | 4.1 | 122 | -0.21 | 0.02 | -0.21 | 0.02 | -0.12 | 0.2 | -0.19 | 0.04 | -0.16 | 0.1 |
| Weather | Weather | 1.5 | 38 | 0.06 | 0.7 | -0.22 | 0.2 | 0.22 | 0.2 | 0.16 | 0.3 | 0.05 | 0.8 |

### 5.2.2 Relation in terms of diurnal usage data

*Higher Books use in the evening, Productivity use in the afternoon and night relate with higher CGPA.* Our findings show that the students who use *Books* category apps more during study time (i.e., evening time) are more likely to have higher CGPA (Duration: $r_s$=0.2, p=0.07; Launch: $r_s$=0.23, p=0.04) (Table 4). In the *Productivity* category also, we find that the students who use apps of this category more during night time (Duration: $r_s$=0.27, p=0.04) and afternoon time (Duration: $r_s$=0.24, p=0.01; Launch: $r_s$=0.23, p=0.01, number of unique apps: rs=0.27, p=0.004), are also more likely to have higher CGPA.

*Higher Video app usage in the morning has a relation with lower CGPA.* The *Video* category shows a statistically significant negative correlation both in terms of usage duration ($r_s$=-0.25, p=0.01) and duration per app ($r_s$=-0.2, p=0.046) of the morning period (Table 4). As aggregated usage data of this category shows a negative relation with academic performance (Section 5.2.1), we were eager to see its relation during study time (i.e., evening time). Interestingly, we find that the number of used apps of the *Video* category during study time ($r_s$=-0.25, p=0.004) negatively correlates with academic performance. Moreover, after considering the number of apps used during sleeping time (i.e., night time), we also find a similar negative association of the *Video* category with academic performance ($r_s$=-0.35, p=0.0002) (Table 4).

*Social Media use shows no significant relation with CGPA in almost every period.* We find that there is a significant negative association between academic performance and the number of used unique *Social Media* apps in the night time range ($r_s$=-0.19, p=0.02) (Table 4). However, similar to the total usage data-based correlation (shown in Table 3), a time range-based analysis of usage duration, frequency of launch, and duration per *Social Media* app also presents that in every period, there is no significant (p>0.05) association (Table 4). In fact, in terms of the number of used unique apps, we do not find any significant association in the other three-time ranges (morning, afternoon, and evening).

Table 4: Correlation of CGPA with four categories usage data in periods. Dark and light grey colored cells present significant and close to significant results respectively.

| App Category | Period | Sample Size (N) | Usage Duration | | Freq. of Launch | | # of Apps | | Duration per App | |
|---|---|---|---|---|---|---|---|---|---|---|
| | | | Coef. ($r_s$) | p | Coef. ($r_s$) | p | Coef. ($r_s$) | p | Coef. ($r_s$) | p |
| Books | Night | 30 | 0.1 | 0.29 | 0.01 | 0.49 | 0.25 | 0.09 | 0.08 | 0.33 |
| | Morning | 46 | -0.01 | 0.46 | 0.09 | 0.29 | 0.03 | 0.42 | -0.02 | 0.45 |
| | Afternoon | 54 | 0.05 | 0.36 | -0.02 | 0.44 | 0.14 | 0.16 | 0.02 | 0.45 |
| | Evening | 57 | 0.2 | 0.07 | 0.23 | 0.04 | -0.07 | 0.29 | 0.23 | 0.04 |
| Productivity | Night | 41 | 0.27 | 0.04 | 0.24 | 0.07 | -0.13 | 0.22 | 0.3 | 0.03 |
| | Morning | 97 | 0.16 | 0.06 | 0.13 | 0.1 | 0.1 | 0.17 | 0.15 | 0.07 |
| | Afternoon | 92 | 0.24 | 0.01 | 0.23 | 0.01 | 0.27 | 0.004 | 0.11 | 0.14 |
| | Evening | 87 | 0.1 | 0.19 | 0.09 | 0.21 | 0.02 | 0.43 | 0.09 | 0.21 |
| Video | Night | 101 | 0.05 | 0.3 | 0.004 | 0.48 | -0.35 | 0.0002 | 0.13 | 0.21 |
| | Morning | 95 | -0.25 | 0.01 | -0.15 | 0.08 | -0.14 | 0.09 | -0.2 | 0.046 |
| | Afternoon | 111 | -0.05 | 0.29 | -0.14 | 0.07 | -0.08 | 0.19 | -0.04 | 0.71 |
| | Evening | 114 | -0.14 | 0.07 | -0.11 | 0.13 | -0.25 | 0.004 | -0.07 | 0.44 |
| Social Media | Night | 109 | -0.08 | 0.2 | -0.11 | 0.14 | -0.19 | 0.02 | 0.05 | 0.61 |
| | Morning | 114 | 0.08 | 0.19 | 0.07 | 0.23 | -0.02 | 0.43 | 0.09 | 0.32 |
| | Afternoon | 113 | -0.01 | 0.44 | -0.07 | 0.22 | -0.1 | 0.15 | 0.11 | 0.24 |
| | Evening | 115 | 0.03 | 0.37 | -0.02 | 0.42 | -0.12 | 0.11 | 0.16 | 0.09 |

## 5.3 Difference of Academic Performance Between High and Low Users

### 5.3.1 Difference in terms of aggregated smartphone usage data

Table 5: Difference of CGPA between the high and low users. Here, the total number of sessions denotes the sessions regardless of the usage duration (in a session). The grey colored cells present the significant (p<0.05) results.

| Total Smartphone Usage Data | Sample Size (N) | High Users CGPA Mean (SD) | Low Users CGPA Mean (SD) | Stats (U) | p | Total Smartphone Usage Data | Sample Size (N) | High Users CGPA Mean (SD) | Low Users CGPA Mean (SD) | Stats (U) | p | Total Smartphone Usage Data | Sample Size (N) | High Users CGPA Mean (SD) | Low Users CGPA Mean (SD) | Stats (U) | p |
|---|---|---|---|---|---|---|---|---|---|---|---|---|---|---|---|---|---|
| Duration | G1: 41 G2: 41 | 3.02 (0.59) | 3.06 (0.78) | 749.5 | 0.5 | # of App | G1: 41 G2: 41 | 2.94 (0.68) | 3.15 (0.595) | 1459 | 0.28 | Micro Session | G1: 41 G2: 41 | 2.96 (0.62) | 3.14 (0.63) | 1304 | 0.04 |
| Launch | G1: 41 G2: 41 | 2.97 (0.66) | 3.19 (0.62) | 670.0 | 0.275 | Duration /App | G1: 41 G2: 41 | 3.08 (0.57) | 3.06 (0.67) | 1643 | 0.76 | Review Session | G1: 41 G2: 41 | 2.94 (0.6) | 3.15 (0.64) | 1243 | 0.038 |
| Duration/ Launch | G1: 41 G2: 41 | 3.09 (0.54) | 2.92 (0.72) | 935.5 | 0.3 | Total # of Session | G1: 41 G2: 41 | 2.82 (0.68) | 3.21 (0.57) | 1031 | 0.002 | Engage Session | G1: 41 G2: 41 | 2.92 (0.69) | 3.16 (0.59) | 1241.5 | 0.04 |

*Users having a higher number of app usage sessions have lower CGPA.* Supporting the findings of correlation analysis, a comparative study also shows students having a higher number of app usage sessions, have significantly lower CGPA (high users' CGPA of 2.82 vs low users' CGPA of 3.21, p=0.002) (Table 5). After dividing the sessions in terms of usage duration (Section 4.2), we also find high users in terms of the number of the micro (high users' CGPA 2.96 vs low users'

CGPA 3.14, p=0.04), review (high users' CGPA 2.94 vs low users' CGPA 3.15, p=0.038), and engage sessions (high users' CGPA 2.92 vs low users' CGPA 3.16, p=0.04) have significantly lower CGPA. On the other hand, in usage duration and launch, low users have relatively higher CGPA than high users (Table 5). But these differences are not statistically significant. This reveals app usage session-based data can be a more important feature in predicting the students' CGPA.

### 5.3.2 Difference after considering all users of each app category

*High users of Books and Productivity have higher CGPA.* Under consideration of all users of the *Books* category, our results show that both in terms of usage duration and launch, high users have statistically significantly higher CGPA than the low users (Duration: high users' CGPA 3.17 vs low users' CGPA 2.86, p=0.038; Launch: high checkers' CGPA 3.27 vs Low checkers' CGPA 2.81, p=0.008) (Table 6). Similar to the *Books* category, the *Productivity* category's high users also have significantly higher CGPA than the low users (Duration: p=0.0194; Launch: p=0.0194) (Table 6). Aligning with findings of correlation analysis, these imply *Books* and *Productivity* categories have a significant positive relation with CGPA.

Table 6: CGPA difference between high and low users, in terms of different data of app categories. Dark and light grey colored cells present significant and close to significant results respectively.

| Usage Data | Category | Sample Size (N) | High and low users' CGPA difference | | | Category | Sample Size (N) | High and low users' CGPA difference | | |
|---|---|---|---|---|---|---|---|---|---|---|
| | | | G1: High Users' CGPA | G2: Low Users' CGPA | p | | | G1: High Users' CGPA | G2: Low Users' CGPA | p |
| | | | Mean (SD) | Mean (SD) | | | | Mean (SD) | Mean (SD) | |
| Duration | Books | G1: 28, G2: 28 | 3.17 (0.48) | 2.86 (0.61) | 0.038 | Video | G1: 40, G2: 40 | 3.07 (0.52) | 3.26 (0.5) | 0.067 |
| Launch | | G1: 27, G2: 24 | 3.27 (0.41) | 2.81 (0.64) | 0.008 | | G1: 40, G2: 40 | 2.99 (0.68) | 3.27 (0.47) | 0.067 |
| # of App | | G1: 28, G2: 55 | 3.18 (0.52) | 3 (0.67) | 0.3 | | G1: 69, G2: 53 | 2.98 (0.67) | 3.18 (0.59) | 0.067 |
| Duration/App | | G1: 28, G2: 55 | 3.18 (0.46) | 3.01 (0.69) | 0.3 | | G1: 40, G2: 82 | 3.01 (0.6) | 3.09 (0.66) | 0.33 |
| Duration | Productivity | G1: 41, G2: 41 | 3.22 (0.54) | 2.92 (0.6) | 0.0194 | Social Media | G1: 39, G2: 39 | 3.03 (0.58) | 2.89 (0.75) | 0.98 |
| Launch | | G1: 41, G2: 37 | 3.26 (0.56) | 2.93 (0.61) | 0.0194 | | G1: 39, G2: 39 | 3.00 (0.6) | 2.99 (0.64) | 0.98 |
| # of App | | G1: 59, G2: 65 | 3.11 (0.66) | 3.04 (0.61) | 0.33 | | G1: 41, G2: 77 | 3.11 (0.51) | 3.03 (0.65) | 0.98 |
| Duration/App | | G1: 41, G2: 83 | 3.21 (0.53) | 3 (0.67) | 0.057 | | G1: 39, G2: 79 | 3.13 (0.62) | 3.03 (0.65) | 0.98 |

*High users of Video apps tend to have lower CGPA while high and low users of Social Media have indifferent CGPA.* In the *Video* category, our findings show that high users have lower CGPA than low users and p-value is closer to the significance level of 0.05 (Table 6). We find such negative relation when we consider high users in terms of *Video* apps' usage duration (high users' CGPA 3.07 vs low users' CGPA 3.26, p=0.067) as well as in terms of launch of *Video* apps (high checkers' CGPA 2.99 vs low checkers' CGPA 3.27, p=0.067). On the other hand, like the previous analyses (Section 5.2) regarding *Social Media*, this analysis also reveals no significant relation of this app category with academic performance. As shown in Table 6, we find no significant (p=0.98) difference in CGPA between high and low *Social Media* users (Duration: p=0.98; Launch: p=0.98, number of used apps: p=0.98, duration per app: p=0.98).

### 5.3.3 Difference after considering only the users having similar smartphone usage behavior

To investigate more about the CGPA difference between the high and low users, we have clustered the users who have similar smartphone usage behavior in terms of total smartphone usage duration and launch. For clustering purposes, we have used the DBSCAN algorithm (Section 4.3.2). To find out the optimal hyper-parameter value of ε (eps), we have followed a previous study [41]. Figure 3 shows that cluster 2 and cluster 3 have a small number of participants (N<12) which makes it difficult to categorize the users. Thus, we have used only the largest cluster (Cluster 1) for this analysis which consists of 80 participants. After that, we grouped the high and low users of each app category to explore what happens to CGPA when users use smartphones for a similar amount but have variations in usage data of the app categories.

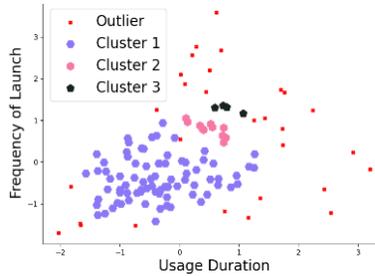

Figure 3: Cluster 1, cluster 2, and cluster 3 represent the students having similar smartphone usage behaviour.

Our analysis shows despite using smartphones for a similar amount, high users of the *Productivity* category have significantly higher CGPA, but high users of the *Video* category have significantly lower CGPA (Table 7). Both in terms of usage duration and frequency of launch, students who spent and launched *Productivity* category apps more, have also significantly higher CGPA than the low users (Duration: high users' CGPA 3.3 vs low users' CGPA 2.98, p=0.03; Launch: high checkers' CGPA 3.37 vs low checkers' CGPA 3.01, p=0.02). On the other hand, students who spent and launched *Video* apps more (i.e., high users of the *Video* category), have significantly lower CGPA (Duration: p=0.025; Launch: p=0.02).

Table 7: High and low users' CGPA difference, when users use smartphones for similar amounts but have differences in app categories usage data. Grey-colored cells present significant results.

| App Category | Data | Sample Size (N) | G1: High Users' CGPA | | G2: Low Users' CGPA | | Test Stats | p |
|---|---|---|---|---|---|---|---|---|
| | | | Mean | SD | Mean | SD | | |
| Books | Duration | G1: 27, G2: 32 | 3.24 | 0.59 | 3.2 | 0.65 | U=440 | 0.45 |
| | Launch | G1: 26, G2: 32 | 3.21 | 0.55 | 3.2 | 0.65 | U=396.5 | 0.45 |
| Productivity | Duration | G1: 27, G2: 27 | 3.3 | 0.57 | 2.98 | 0.68 | U=473 | 0.03 |
| | Launch | G1: 26, G2: 27 | 3.37 | 0.58 | 3.01 | 0.66 | U=483 | 0.02 |
| Video | Duration | G1: 27, G2: 27 | 2.98 | 0.68 | 3.34 | 0.53 | U=251 | 0.025 |
| | Launch | G1: 27, G2: 26 | 2.96 | 0.77 | 3.4 | 0.39 | U=224.5 | 0.02 |
| Social Media | Duration | G1: 27, G2: 27 | 3.09 | 0.65 | 2.94 | 0.85 | U=395 | 0.6 |
| | Launch | G1: 27, G2: 27 | 2.99 | 0.67 | 3.12 | 0.62 | t(54)=-0.78 | 0.6 |

## 5.4 High and Low CGPA Holders' Usage Data Difference

### 5.4.1 Difference in aggregated usage data

*High CGPA holders use Productivity apps more.* In the *Productivity* category, we find high CGPA holders spend (p=0.0056) and launch (p=0.0096) significantly higher number of times. In addition, they have a higher per-app duration (p=0.004). In the *Books* category also, we find a higher usage duration, launch, and per-app duration of the high CGPA holders, though the p-value (0.066) is slightly higher than the significance level of 0.05.

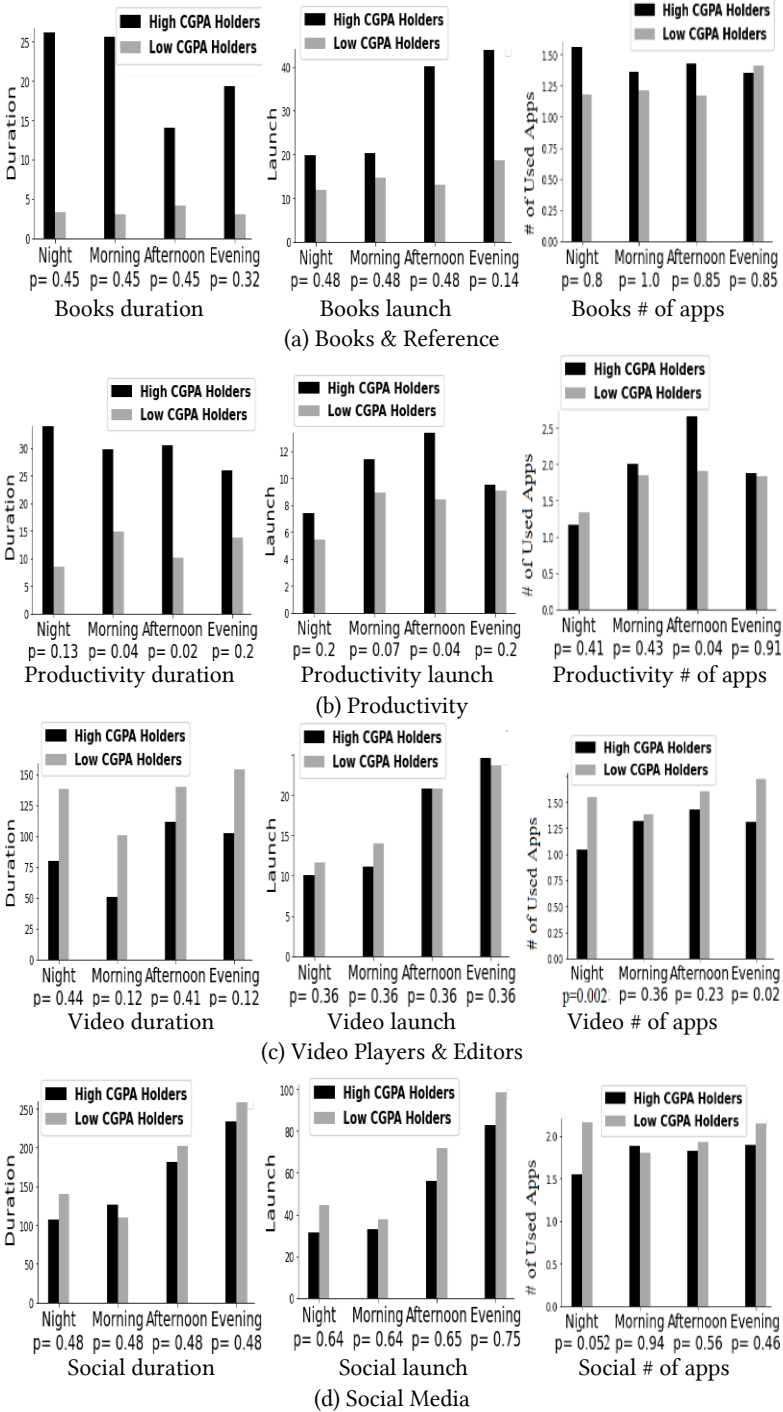

Figure 4: Diurnal usage pattern of the high and low CGPA holders in case of (a) Books & Reference, (b) Productivity, (c) Video Players & Editors, and (d) Social Media category.

*High CGPA holders use Video apps less.* In all previous analyses (Section 5.2 and Section 5.3), we find a significant negative relation of the Video category's app usage with CGPA. Thus, we expect high CGPA holders will use this category's apps less than low CGPA holders. Expectedly, we find high CGPA holders spend (high CGPA holders: 286.79 minutes vs low CGPA holders: 500.91 minutes, p=0.044) and launch (p=0.048) significantly less (Table 8). Also, they use significantly fewer Video apps (p=0.048) which may indicate that they keep concentration on a few apps in this category.

Table 8: High and low CGPA holders' difference in usage data of app categories. Duration is in minutes. Cat.: Category. Dark and light grey colored cells present significant and close to significant results respectively.

| Usage Data | Cat. | Sample Size (N) | G1: High CGPA Holders | G2: Low CGPA Holders | p | Cat. | Sample Size (N) | G1: High CGPA Holders | G2: Low CGPA Holders | p |
|---|---|---|---|---|---|---|---|---|---|---|
| | | | Mean (SD) | Mean (SD) | | | | Mean (SD) | Mean (SD) | |
| Duration | Books | G1: 21, G2: 29 | 49.62(151.48) | 7.09 (7.85) | 0.066 | Video | G1: 32, G2: 46 | 286.79(261.44) | 500.91(491.47) | 0.044 |
| Launch | | G1: 21, G2: 29 | 81.24 (283.57) | 30.41 (104.14) | 0.066 | | G1: 32, G2: 46 | 51.84 (59.59) | 60.24 (48.23) | 0.048 |
| Duration/App | | G1: 21, G2: 29 | 19.14 (38.57) | 5.48 (5.97) | 0.066 | | G1: 32, G2: 46 | 214.08 (236.15) | 319.62 (425.45) | 0.09 |
| # of App | | G1: 21, G2: 29 | 1.48 (0.81) | 1.31 (0.66) | 0.45 | | G1: 32, G2: 46 | 1.56 (0.72) | 2.02 (1.06) | 0.048 |
| Duration | Product | G1: 32, G2: 46 | 80.31 (111.8) | 42.57 (98.79) | 0.0056 | Social Media | G1: 29, G2: 43 | 643.28 (256.16) | 682.89 (434.81) | 0.78 |
| Launch | | G1: 32, G2: 46 | 24.06 (16.66) | 20.85 (31.26) | 0.0096 | | G1: 29, G2: 43 | 190.9 (131.54) | 226.7 (187.24) | 0.78 |
| Duration/App | | G1: 32, G2: 46 | 22.47 (33.57) | 10.74 (21.13) | 0.004 | | G1: 29, G2: 43 | 386.44 (244.12) | 319.44 (265.85) | 0.44 |
| # of App | | G1: 32, G2: 46 | 3.53 (1.29) | 3.17 (1.64) | 0.31 | | G1: 29, G2: 43 | 2.1 (1.21) | 2.42 (1.12) | 0.52 |

### 5.4.2 Difference in diurnal usage pattern

*High CGPA holders have significantly different diurnal usage patterns.* We find in case of the *Productivity* category, high CGPA holders spend more time in the morning and afternoon (morning, p=0.04; afternoon, p=0.02) and launch apps significantly a greater number of times than the low CGPA holders in the afternoon (p=0.04) (Figure 4 (b)). Moreover, the high CGPA holders use a higher (p=0.04) number of *Productivity* category apps in the afternoon. On the other hand, high CGPA holders use significantly less number *Video* apps (night, p=0.002; evening, p=0.02) during study time (i.e., evening time) as well as during sleeping time (i.e., night) (Figure 4(c)). On the other hand, in the case of *Social Media*, high and low CGPA holders do not have statistically significant (p>0.05) different usage patterns (Figure 4(d)).

## 5.5 Predicting Students' Academic Performance Using App Usage Data

At first, we used the Lasso and Elastic net algorithms to select the important features, however, these feature selection approaches did not perform well (Table S.4 and Table S.5 of Supplementary). Later, we do correlation analysis for feature selection where the performance of the predictive models deteriorated after adding data of per app duration and duration per launch (Table S.6, Table S.7, and Table S.8 of Supplementary). Interestingly, we find better predictive performance using the core features duration, frequency of launch, and number of used apps which have statistically significant relation with CGPA. Among the 720 features, this approach selected 44 features as important (for details Table S.3 of Supplementary). The models based on these data are not over fitted as presented by the smaller difference in mean absolute error (MAE) value between the training and validation phase (Table 9). The KNN algorithm-based model demonstrates the best performance where the MAE value is 0.36 (Table 9). This says the predicted CGPA is within ±0.36 of the ground truth CGPA.

Beside this, the prediction of Voting regressor and SVR algorithms are also close to the KNN and MAE values are 0.38 and 0.40 respectively (Table 9). To understand how the predicted value of the ML model's (Table 9) relates with the actual CGPA, we also do a correlation analysis. We find that the predicted CGPA of the KNN ($r_s$=0.44, p=0.01), Voting Regression ($r_s$=0.42, p=0.01), and SVR ($r_s$=0.34, p=0.04) significantly positively relate with the actual CGPA (Table 9). This reveals that the higher the actual CGPA, it is more likely to have a higher CGPA in the prediction as well.

Table 9: Performance of the predictive model in 5-fold cross-validation and testing when features are selected by correlation analysis. Dark and light grey colored cells present significant and close to significant results respectively.

| Model Name | Average MAE value in 5-fold cross validation (70% data) | | | Test (30% data) | | Relation of predicted CGPA with actual CGPA | |
|---|---|---|---|---|---|---|---|
| | Sample Size | Avg. MAE in Training | Avg. MAE in Validation | Sample Size | MAE | Correlation Coef. ($r_s$) | p |
| KNN | 84 | 0.38 | 0.46 | 37 | 0.36 | 0.44 | 0.01 |
| Voting regressor | 84 | 0.39 | 0.45 | 37 | 0.38 | 0.42 | 0.01 |
| SVR | 84 | 0.42 | 0.46 | 37 | 0.4 | 0.34 | 0.04 |
| Gradient boosting | 84 | 0.39 | 0.46 | 37 | 0.42 | 0.26 | 0.12 |
| Elastic net | 84 | 0.44 | 0.47 | 37 | 0.43 | 0.27 | 0.11 |
| Random forest | 84 | 0.4 | 0.48 | 37 | 0.43 | 0.17 | 0.32 |
| Lasso | 84 | 0.47 | 0.49 | 37 | 0.43 | 0.19 | 0.25 |
| MLP | 84 | 0.55 | 0.56 | 37 | 0.5 | -0.03 | 0.87 |
| XGBoosting | 84 | 0.9 | 0.95 | 37 | 1.02 | 0.3 | 0.07 |

## 6 DISCUSSION

Our analysis demonstrates smartphones' aggregated usage data such as duration which was much explored (e.g., in [5, 6]) do not have a statistically significant relation. Instead, we find the number of app usage sessions has a significant negative relation with academic performance. We also find some app categories such as *Books & Reference* and *Productivity* have a positive relation. These findings extend the studies [13, 22, 23, 24, 25, 34] which present the relation of these app categories (i.e., *Video, Books, Productivity*) using self-reported data and also found only the negative relation with smartphones on academic performance [5, 6]. Apart from these, our analysis demonstrates the varying app categories usage behavior of the high and low CGPA holders. Extending the previous studies [15, 48, 73, 81], we present a faster and minimalistic approach to predict academic performance instantly which can be potential for the students, parents, and teachers for estimating academic performance and taking early intervention.

### 6.1 App Usage Sessions Negatively Relate to Academic Performance

We find the people who have a higher number of app usage sessions are more likely to have lower academic performance. Moreover, the number of micro and review sessions also has a significant negative relation with academic performance. Grace-Martin and Gay [50] found a negative relation of session length in the case of *Browser* apps and our empirical research extends the previous study presenting negative relation in the case of total smartphone usage data. We also find a negative relation of micro and review sessions if that usage occurs in the afternoon and evening which is the class and study time respectively in Bangladesh. One plausible reason for having such a negative relation is students' daily number of sessions positively relates to sessions during the class time [11] and they get distracted by smartphones during the class time [11, 80]. Meanwhile, in our study, micro and review session users spent a maximum of 15 seconds and 60 seconds respectively which is a very short time. Therefore, having a higher number of sessions indicates they use smartphones more after a time interval (i.e., more than 45 seconds in our study) and this can distract the students from their study as different app usage creates different types of emotions [94]. Besides, accessing phones a higher number of times to spend less time can present high session users' multitasking which has a negative impact on learning [75] and academic performance [13, 75]. Also, multitasking creates stress on the students [28]. Thus, mitigating higher app usage sessions can be beneficial for the students.

## 6.2 Different App Categories Relate to Academic Performance Differently

***Books & Reference and Productivity apps relate with CGPA positively.*** It was interesting to investigate the relation of *Productivity* and *Books & Reference* categories with students' academic performance as these app categories are more popular than *Social Media* [36, 37] and were unexplored in previous studies using objective data. We find usage data (e.g., duration) of these two app categories significantly correlate with academic performance. High users have significantly higher CGPA than low users. In fact, when we cluster the students who have similar smartphone usage behavior, we find that still, the high users of the *Productivity* category have statistically significantly higher academic performance. The positive relation can be due to the fact that apps of this category help students in different ways. For instance, *Productivity* apps such as *Google Drive, Document Viewer, Notes, Calendar*, etc. help students in learning [22] and study duration is positively correlated with academic performance [15]. Moreover, *Books & Reference* category's apps such as the dictionary, and grammar-related apps help students in learning [23, 25] and comics apps help them in reducing stress and boosting their thinking skills [24]. Nowadays, both teachers [80] and parents [57] are concerned about the technology usage of the students and parents set different rules to reduce the technology usage of their children [57]. Our findings suggest that smartphones do not always negatively relate and it can support students' learning needs greatly depending on students' usage of app categories. These findings can be insightful for parents who perceive smartphones as detrimental to their children's education.

***Social Media category does not have a significant relation with CGPA.*** Our correlation and comparative study show there is no significant relation of *Social Media* with academic performance though Dhaka is the second most active *Facebook* users' city [39]. Our findings are consistent with several prior research which also do not find any significant difference in academic performance between high and low checkers of *Facebook* [1], *Instagram* [47], and *Social Media* [26] apps. We also find that the low CGPA holders do not use *Social Media* significantly more during study time as in other time periods of a day. This finding can explain the plausible reason for not having any significant relation of this app category on students' academic performance. On the other hand, high *Facebook* checkers spend less time in each session of *Facebook* use which can present their self-regulation behavior [1]. Apart from these, depending on the purpose of using *Social Media*, negative influence can be mitigated. For instance, *Facebook* usage for general purposes does not have any significant relation with academic performance [3]. Therefore, these findings again highlight that app usage does not always negatively relate to academic performance.

***Video Players & Editors category has a negative relation with CGPA.*** We find a negative correlation between students' academic performance and *Video Players & Editors* category's usage data. In addition, we find during the study time, low CGPA holders spend significantly more time on this category's apps. This can be the cause of having a negative impact on low CGPA holders as *Video Players & Editors* related apps (e.g., *YouTube* [34]) distract the students. Moreover, watching videos and media files on devices is negatively correlated with academic performance [13]. On the other hand, previous work [19] shows people watch videos to be inspired most often in the morning. It was surprising to see that our findings show a significant negative relation between morning time *Video Players & Editors* category apps use and academic performance. It can be due to the fact that in participants' universities, classes start at 9.00 AM, and in the morning, low CGPA holders spend significantly more time on apps of this category. This reveals that at that time, they may prefer to watch videos, instead of concentrating or participating in classes. Watching videos on *YouTube* can cause problems such as failing to keep track of time [28]. Therefore, students' study during class time can be hampered. Meanwhile, apps such as *YouTube* which are entertaining can be more distractful as the effect of the entertainment apps on joy and emotion is stronger [94]. Low CGPA holders of our research used a significantly higher number of *Video* apps during the study time (6 PM to 12 AM) as well as the sleep time (12:01 AM to 6 AM). This points out that *Video* apps can distract them from their study. In addition, late-nighters tend to multitask more [28] which has a negative impact on academic performance [13]. Therefore, self-regulation through context (e.g., study time, class time) aware intervention systems can help the students to concentrate more on the study and mitigate the negative influence of the *Video Players & Editors* app category.

## 6.3 Design Implications

***Intervention regarding app usage sessions.*** Through our statistical analysis, it is found app usage sessions relate to academic performance negatively. As we discussed in Section 6.1, it indicates the multitasking preferences of the students. However, people are unaware of their multitasking capabilities [82] and multitaskers also have less capability to filter irrelevant content [86]. Therefore, the designed system can set an input task-based interruption [69] which can remind the user about the necessity of using an app and thus, it can filter the less important usage which in turn may reduce the number of usage sessions. The system can nudge (e.g., through playing a vibration [83]) the user when the number of app usage sessions becomes higher in a shorter time. Raising consciousness about the type of app usage sessions through that system can be beneficial since we find micro and review sessions relate to academic performance significantly negatively. In addition, by becoming aware of the contexts, the system can take more steps (e.g., intervention through a higher input task [69]) to discourage frequent app usage during class time and study time. However, as both the teachers and students prefer systems to have students' autonomy [80], by respecting it, the system should have the capability to incorporate the student's self-defined rules.

***Intervention for Video Players & Editors, Books & Reference, and Productivity apps' usage.*** We find that the *Books & Reference* and *Productivity* app categories positively relate to students' academic performance. Therefore, the system can encourage the students to use the apps of *Books & Reference* and *Productivity* category more in groups as the researchers [71, 84] found group-based intervention performs better in reducing smartphone overuse and encouraging them to concentrate more on the study. For example, when one student studies through an app of *Books & Reference* or *Productivity* category, she can share it with her classmates to study together by limiting phone usage. This approach can result in longer study duration as the group-based study is enjoyable and peer-influence work there [84]. On the other hand, the system can discourage the usage of the *Video Players & Editors* app category where attractive entertaining content may distract the students. However, as videos from apps such as *YouTube* can also be watched for information or to understand a concept [19], some steps such as goal reminder-based intervention [61] can be used to avoid distraction from an app. In addition, the system must have to understand the context of app usage as goal prompts are annoying to some students [61]. Meanwhile, previous studies suggest filtering (e.g., filtering notifications [79]) options to get less distracted. Thus, while using apps such as *YouTube*, filtering videos on the basis of students' preferences (e.g., in the sidebar, showing only the videos related to the study topic) can help the students to be more focused on their watching goal.

***Minimal system to predict CGPA faster:*** Using only app usage data, we develop ML models which can predict academic performance accurately. However, we find that predictive models do not perform well when the features are selected using Elastic net and Lasso algorithms. One plausible explanation behind this is that most of the feature selection methods do not perform well when the number of features becomes higher than the number of samples [56]. On the other hand, when we selected the features through correlation analysis, KNN predicted CGPA accurately and the predicted CGPA was within ±0.36 of the actual CGPA. Though a previous study [15] predicted students' CGPA within ±0.17 of the actual CGPA, their ML models relied on semester-long sensing as well as self-reported data. Our presented ML model used only the past 7 days' app usage data which can be retrieved in less than a second without users' intervention in the runtime and without running in the background. Moreover, the previous study's data need additional preprocessing which can be computationally and power expensive (covered in Section 3) whereas our supervised ML models rely on simple mathematical calculation-based app usage data (covered in Section 4.2). Therefore, our faster, unobtrusive, and minimalistic approach can be supportive for the students, teachers, and parents for instantly estimating academic performance and taking early intervention based on that. This also creates new opportunities for the HCI designers to design a system that can collect app usage data with students' proper consent and predict their grades in a minimalistic and faster approach.

## 6.4 Ethical Implications

To predict academic results, unavailability (e.g., in Bangladeshi universities) of technology such as smart cards [81, 89] and unaffordability [101] of costly technology such as Fitbit [100] can make such technology-based systems infeasible in low and middle-income countries (LMICs) which in turn can bias the benefits of pervasive technology and increase disparities. However, we developed a system based on smartphones which are available to a large number of students of LMICs (e.g., 86.62% of university students in Bangladesh use smartphones [102]). In addition, we took several steps to overcome some limitations of LMICs. For instance, unlike other systems [15], our system does not need to run in the background and also does not rely on a power-consuming GPS sensor [103] which could significantly consume the battery power [103] creating a barrier [104] in the applicability in low-resource settings. The resource-efficient steps can make our minimal system applicable in LMICs and may contribute to decreasing the disparities and inequalities in educational technology facilities between the LMICs and developed countries. We believe and argue that the system designers have an ethical imperative to make the system applicable in LMICs since approximately 90% of youth of the world reside in LMICs [109] and there persists a large gap in having advanced technology [110].

In South Asian countries [107] including Bangladesh [108], students having lower academic grades are looked at negatively and students remain under pressure. Consequently, we caution our work against being misused if students who are identified as lower grade holders are viewed down or are forced to have higher grades. We contend that the parents, caregivers, and teachers should behave with the low-grade holders with empathy and support to maximize the student's academic success using our system. Besides, steps for intervention or other applications of our system such as course advising should not be solely based on the predicted grades since qualitative information is not incorporated in a grade prediction tool [113] which may have a negative impact. Accordingly, human judgment and also students' aims, family, and health circumstances should be considered [113] for taking a step toward students' improvement.

As for the ML models' features, we used data from the apps of different categories which can facilitate the researchers to develop a system by respecting one's personal preferences since all students may not have preferences to allow a system to retrieve data of all app categories as found in previous research [105] in mental health prediction. Besides, we found high CGPA holders have significantly different behavioral patterns than low CGPA holders. As a result, a personalized system can be more impactful which can reduce the biases towards the general students and perform well for the students as well whose behavior deviates from others. Additionally, such a personalized system can play a role in mitigating students' concerns [105] about the loss of control, dignity, and autonomy. Along with personalizing the system, the federated learning technique [106] can also be incorporated where data will remain on the users' smartphones and will not get processed on the server side which can enable the system to have better privacy.

## 7    LIMITATIONS AND FUTURE WORK

Inspired by the previous HCI research [15, 32, 88, 90], we did correlation analyses to find the relation of CGPA with different app categories. To answer the same research question, we did comparative studies in three different ways. Interestingly, all of these analyses revealed almost similar findings. However, still, these may not present a causal analysis. Moreover, since some app categories are associated with CGPA, further studies can explore the app-level analysis of those categories where our research can work as a precursor. Besides, we believe that due to being minimal and very faster (<1 second), our tool uses a very less amount of resources (e.g., consumption of battery power, bandwidth) of a phone. But in our future study, we will explore to get precise information about the used amount of resources by our tool.

During the COVID-19 pandemic, apps such as *YouTube* usage has been increased significantly [99] which may have an impact on academic grades as we found a significant negative relation with *Video Players & Editors* category apps. Thus, there is a need for an in-depth investigation to find the subtle changes in app usage behavior amid the pandemic and its link with grades. We believe that our promising findings will inspire the researchers to explore more in the context of the pandemic where our study can be seen as the precursor. As a first step, we have already constructed a dataset from January to June 2021 where in the last week of March 2021 Bangladesh experienced the second wave of COVID-19 [112]. In that study, we collected 105 students' CGPA and app usage data where the participants were from 8 different higher educational institutes and 51.6% districts of Bangladesh. In the future, we will analyze the dataset overcoming the aforementioned limitations, and share the promising findings with the scholarly community.

## 8    CONCLUSION

We have presented an in-the-wild study regarding the relation of smartphone usage data with academic performance. Analyzing 124 students' 7 days' data, we have come up with different app usage behavior along with their relation with CGPA. Our analysis also presents high and low CGPA holders differ by aggregated data and also by diurnal pattern (p<0.05). Based on our instantly (Median=0.21s, Mean=0.31s, SD=1.1s) accessed app usage data, we predict CGPA where our KNN model's predicted CGPA is within ±0.36 of the actual CGPA. Thus, the research insights of our study are expected to add value to learners, caregivers, and educators about technology usage and its impacts. In addition, our presented approach having a minimalistic intervention to predict academic performance can be used for early intervention and also in the concrete design process to be aware of app usage choices, making possible positive academic outcomes.

### ACKNOWLEDGMENTS

We thank the participants for their time in data donation. In addition, we thank Md Reazul Islam and Tushar Asad for their invaluable support to collect data. Besides, we thank the faculties of Eastern University Bangladesh, particularly, Mohammad Rifat Arefin, Sanjana Srabanti, Prof. Md. Abbas Ali Khan, and Prof. Md. Mahfuzur Rahman who cordially supported this research in various ways.